\documentclass[aps,prb,preprint,superscriptaddress]{revtex4-1}
\usepackage{subfig}
\usepackage{graphicx}

\newcommand{\VO}{V$_{\mathrm O}$}

\begin{document}


\title{How the aggregation of oxygen vacancies in rutile based TiO$_{2-\delta}$ phases causes memristive behavior.}


\author{Wolfgang Heckel}
\email[]{wolfgang.heckel@tuhh.de}
\affiliation{Institute of Advanced Ceramics, Hamburg University of Technology, D-21073 Hamburg, Germany}

\author{Michael Wehlau}
\affiliation{Bremen Center for Computational Materials Science, University of Bremen, D-28359 Bremen, Germany}

\author{Sascha B. Maisel}
\affiliation{Institute of Advanced Ceramics, Hamburg University of Technology, D-21073 Hamburg, Germany}

\author{Thomas Frauenheim}
\affiliation{Bremen Center for Computational Materials Science, University of Bremen, D-28359 Bremen, Germany}

\author{Jan M. Knaup}
\altaffiliation{now: Atotech Deutschland GmbH, Berlin, Germany}
\affiliation{Bremen Center for Computational Materials Science, University of Bremen, D-28359 Bremen, Germany}

\author{Stefan M\"uller}
\affiliation{Institute of Advanced Ceramics, Hamburg University of Technology, D-21073 Hamburg, Germany}


\date{\today}

\begin{abstract}
The results of a comprehensive and systematic ab-initio based ground-state
search for the structural arrangement of oxygen vacancies in rutile
phase TiO$_2$ provide new insights into their memristive properties. We find that
O vacancies tend to form planar arrangements which relax into structures exhibiting metallic
behavior. 
These meta-stable arrangements are structurally akin to, yet distinguishable from the Magn\'eli phase. They
exhibit a more pronounced metallic nature, but are energetically less favorable. Our results 
confirm a clear structure-property relationship
between segregated oxygen vacancy arrangement and metallic behavior in reduced oxides.\end{abstract}

\pacs{85.30.-z,71.15.Mb,71.20.-b,72.20.-i}
\keywords{Memristor, vacancy ordering, reduced TiO$_2$, Magn\'eli phase}

\maketitle

\section{\label{sec:intro}Introduction}
Understanding the behavior of O vacancy defects (V$_{\mathrm O}$) 
in reducible 
metal oxides is of paramount importance to understanding a wide range of
highly interesting physical and chemical effects. These include oxide catalysis for
sustainable CO$_2$ conversion into chemical fuels, highly efficient batteries for
sustainable energy storage and electromobility, and innovative electronic devices with
the potential to fundamentally transform information technology.
The latter is expected to be enabled by the memristor,
a fascinating new electronic device that complements the
widely known passive two-terminal devices of resistor, capacitor and 
inductance~\cite{chua_memristor_1971,chua_resistance_2011}.
It acts as an ohmic resistance depending on the history of the current which
has flown through it in the past---hence the name which is a contraction of
memory and resistor. Already proposed in the 1970s by Chua~\cite{chua_memristor_1971}, 
the purposeful
implementation of memristors could only be achieved in 2008~\cite{strukov_missing_2008}. 
Memristors
are being considered for two main applications: First, resistive
switching memory will to scale 
far beyond current FLASH memory
technology~\cite{waser_redox-based_2009,clima_rrams_2014}, 
which relies on tunneling electrons to and from floating gates in CMOS transistors. 
From this arises the possibility to shift 
computing paradigms away from the current
compute centric Von Neumann architecture towards storage centric concepts~\cite{strukov_hybrid_2010}
which avoid data transfer bottlenecks in the former.
Second, the memristor forms an indispensable building block of electronic analogues
to nerve cells for neuromorphic computing~\cite{jo_nanoscale_2010,wang_synaptic_2012}. 
Without it, modeling synaptic behavior requires
extremely complex active electronics. This complexity severely limited the possible
number of interlinks between artificial neurons, rendering the construction of powerful neuromorphic devices infeasible.
Recently, an artificial neuronal network was implemented in hardware, interlinking
576 artificial neurons in Si based CMOS by a matrix of 73728 artificial 
synapses realized
by WO
memristors on the same chip~\cite{cruz-albrecht_scalable_2013}. The device exhibits considerable
learning ability and constitutes a significant advance in neuromorphic computing.
At the same time, enormous
effort is exerted in the development of 
resistive switching memory based on transition metal oxide memristors. Laboratory
devices already surpass speed and durability of FLASH memory cells~\cite{clima_rrams_2014}.

The first material in which memristors were implemented was TiO$_{2-\delta}$ and it is still
an enormously useful and interesting model system~\cite{strukov_missing_2008}. It is understood that the resistance
modulation is based on the migration of oxygen vacancy defects and the formation and
dissolution of conductive~\cite{inglis_electrical_1983} Magn\'eli 
phases~\cite{strachan_structural_2009,kwon_atomic_2010} which are based on planar arrangements
of oxygen vacancies (\VO) in the rutile structure and have Ti$_n$O$_{2n-1}$ 
stoichiometry~\cite{marezio_structural_1973,le_page_structural_1982}. High precision theoretical 
calculations~\cite{park_impact_2011} have shown that 
electrical conductivity can be induced in rutile by different kinds of vacancy 
arrangements, bearing only marginal resemblance to Magn\'eli phases. However, this earlier 
study was limited to a small number of examined vacancy arrangements by the prohibitive
computational cost of ab-initio methods that can describe the electronic structure of
V$_{\mathrm O}$ in TiO$_2$~\cite{deak_polaronic_2011}. Still, combined with the finding that rutile is severely 
destabilized by oxygen deficiency, as evidenced by a pronounced reduction in melting
temperature~\cite{knaup_reduction_2014}, these results lead to an important question for the understanding of the 
memristive effect in transition metal oxides in general: Is the observed formation of
Magn\'eli phases prerequisite for the effect or is the resistance change caused by a 
larger class of structural features, with the Magn\'eli phase being the thermodynamically
most favorable structure exhibiting these features?

With an exhaustive ground-state search, attained by a density functional theory (DFT) based cluster expansion (CE)
approach, we shed light on that question. In this letter, we present the first
systematic investigation of the full configuration space for \VO\ in TiO$_2$
rutile. We identified planar \VO\ arrangements to be energetically relevant and
responsible for the metallic behavior in reduced oxides.

\begin{figure}
\includegraphics[width=8cm]{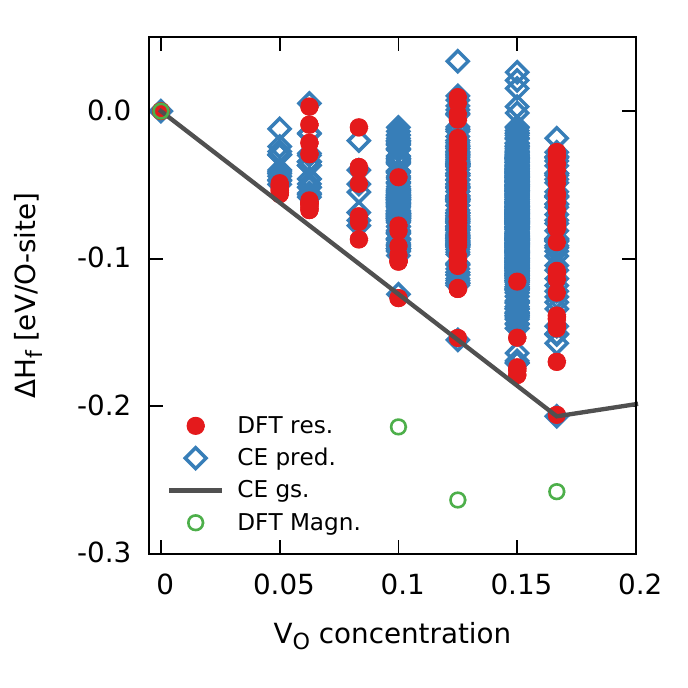}%
\caption{\label{fig:ceenergies} (color online) Formation enthalpies $\Delta
  H_{\mathrm f}$ (see Eq.~\ref{eq:CE-Eform}) of vacancy orderings in the
  rutile crystal structure resulting from a DFT based (filled red circles) CE
  fit (open blue squares). The black solid line marks the CE ground
  states. For comparison, the DFT formation enthalpies of the Magn\'eli phase
  are also displayed (open green circles).
}
\end{figure}

\section{\label{sec:methods}Methods}

In order to explore the full configuration space of \VO\ ordering in reduced TiO$_2$ rutile, we
 initiated a cluster expansion (CE)~\cite{sanchez_generalized_1984}
 using the \textsc{Uncle} code~\cite{lerch_uncle:_2009}
 in the \VO\ concentration range of
 $0\ldots16.7\,\%$. The total energies of the CE input structures were determined by DFT calculations employing
 \textsc{Vasp}~\cite{kresse_efficient_1996,kresse_efficiency_1996}.
 We have applied the PAW-GGA~\cite{blochl_projector_1994,kresse_ultrasoft_1999}
 functional 
 PBEsol~\cite{perdew_restoring_2008}
 with a conjugate gradient algorithm to relax the structures with respect to cell volume, cell shape and 
 atomic positions until the forces on each atom were smaller than $5\,\text{meV}/\text{\AA}$. 
 The energy cutoff of the plane-wave basis was set to $520\,\text{eV}$.
To keep the computational effort within tractable limits, the considered cells
were restrained to every possible unit cell of O defective rutile containing
10 or less Ti atoms. \footnote{The smallest possible symmetric cell was chosen
  for each configuration.} Every such cell predicted to be energetically
favorable by the CE was calculated by DFT (cf.\ Fig.~\ref{fig:ceenergies}).

For the examination of the density of states, additionally, the geometries of
selected structures were relaxed by applying the polaron correction developed
by Lany and
Zunger~\cite{lany_assessment_2008,PhysRevB.80.085202,lany_many-body_2010} and,
subsequently, a static HSE06~\cite{heyd_hybrid_2003,heyd_erratum:_2006}
calculation approximates the electronic ground state.
In their polaron correction, Lany and Zunger introduce the additional potentials
\begin{equation}
V_{\text{hs}} = \lambda_{\text{hs}} \left(1 - n_{m,s} / n_{\text{O,rutile}} \right) \label{eq:LZp}
\end{equation}
for hole states and
\begin{equation}
V_{\text{es}} = \lambda_{\text{es}} \left(1 - p_{m,s} / p_{\text{Ti,rutile}} \right) \label{eq:LZn}
\end{equation}
for electron states to satisfy Koopmans theorem. $n_{m,s}$ and $p_{m,s}$ are
fractional occupancies of state $m$ with spin $s$, $n_{\text{O,rutile}}$ and
$p_{\text{Ti,rutile}}$ denote the occupancies in undisturbed, defect-free
TiO$_2$ rutile, and $\lambda$ defines the coupling. We set these potentials
for O 2p hole states to $\lambda_{\text{hs}} = 4.81\,\text{eV}$ and Ti 3d
electron states to $\lambda_{\text{es}} =
4.18\,\text{eV}$. \footnote{S.\ Lany, private   communication.} Applying the
polaron correction, only subtle differences in defect geometries occur, which
however might have a strong impact on hybrid exchange band structures. 

The Brillouin zone sampling of all structures,
independent of cell shape or size, was performed  using a $7\times7\times7$
Monkhorst−Pack~\cite{monkhorst_special_1976} mesh for PBEsol and Lany-Zunger
corrected calculations, and a $3\times3\times3$ mesh for HSE06 calculations.

The CE Hamiltonian
\begin{equation}
  E(\sigma) = \sum_{C\in\mathbf{C}} J_C\Pi_C(\sigma) \label{eq:CE-Hamiltonian}
\end{equation}
expands the energy of a given structure $\sigma$ into a sum over $n$-body
correlation functions $\Pi_C(\sigma)$ times expansion coefficients $J_C$. A
genetic algorithm elaborated in Refs.~\citenum{hart_gen-algo:2005} and \citenum{Blum2005} is used to
truncate the sum of Eq.~\ref{eq:CE-Hamiltonian} and reduce it to the most relevant $n$-body
interactions. Then, the expansion coefficients $J_C$ of this optimized set $\mathbf{C}$
are extracted by fitting them to a first-principles data base. A
cross-validation scheme~\cite{Stone1974} is used to control the error of the
fits.

The CE Hamiltonian can then be used to evaluate the energetics of
more than 300\,000 structures quickly and efficiently. Whenever the CE
Hamiltonian predicts a structure to be energetically more favourable than the
DFT input structures, it is relaxed with DFT and added to the input set for
the next iteration of the CE. This procedure is reiterated until the CE is
self-consistent. The quality of the fits is observed to vary smoothly with the
number of input structures and the number of non-zero $n$-body interactions.

The final CE yielding the results shown in this paper uses a grand-total of
143 input structures, including the pure elements. The fit incorporates 35
interactions as determined by the genetic algorithm, which can be found in
the supplement.
The Hamiltonian yields a
cross-validation score of $8.9\,\text{meV/O-site}$. This
ensures a sufficient accuracy for the CE energy predictions.

Formation enthalpies in Figure~\ref{fig:ceenergies} are calculated according to
\begin{equation}
  \Delta H_{\text{f}} = E(\sigma) - x \cdot E(\text{100\,\%~\VO}) - (1-x) \cdot E(\text{TiO$_2$~rutile}) \label{eq:CE-Eform},
\end{equation}
where $E(\sigma)$ denotes the total energy of structure $\sigma$ and $x$ the \VO\ concentration of that structure.
$E(\text{TiO$_2$~rutile})$ and $E(\text{100\,\%~\VO})$ correspond to the
perfect stoichiometric TiO$_2$ rutile ($0\,\%$ \VO) and the $100\,\%$ vacancy pseudo-element
concentration energy, respectively.
To identify energetically favorable \VO\ orderings, we assign the total energy of a TiO$_2$ rutile structure
 without any oxygen atoms to $E(\text{100\,\%~\VO})$. This leads to apparent ground states at 
 intermediate vacancy concentrations, as displayed in Fig.~\ref{fig:ceenergies}. As one expects, no ground 
 state between perfect TiO$_2$ rutile ($0\,\%$ \VO) and pure Ti ($100\,\%$ \VO) would appear if 
 the total energy of hcp-Ti were assigned to the $100\,\%$ \VO\ structure.

\begin{figure*}
\begin{centering}
\subfloat[\label{fig:magneli_111}view in {[111]}]{\includegraphics[width=6.5cm]{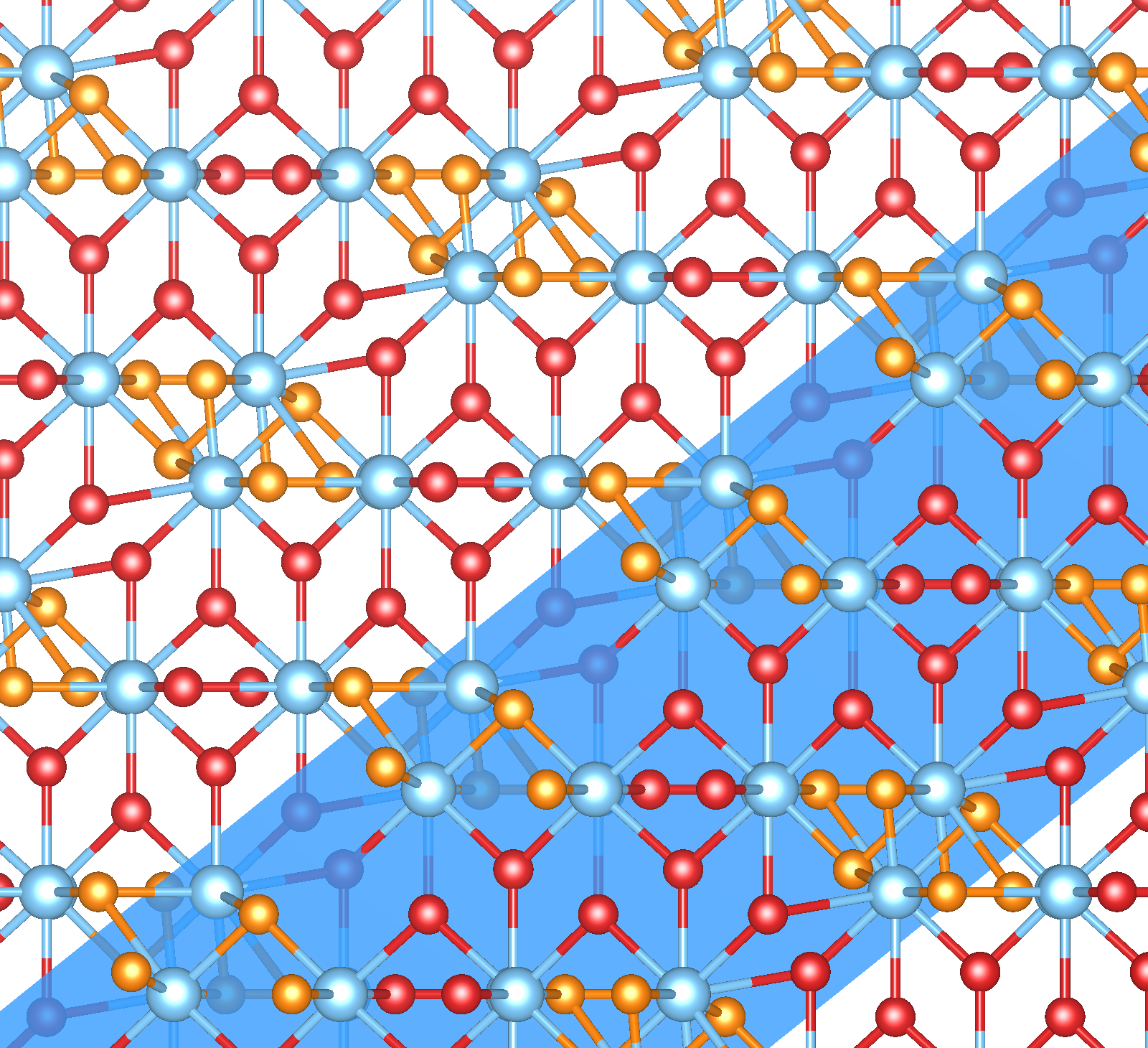}}%
\includegraphics[width=1.0cm]{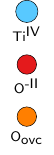}~
\subfloat[\label{fig:magneli_001}(001) view, for clarity one layer only]{\includegraphics[width=6.5cm]{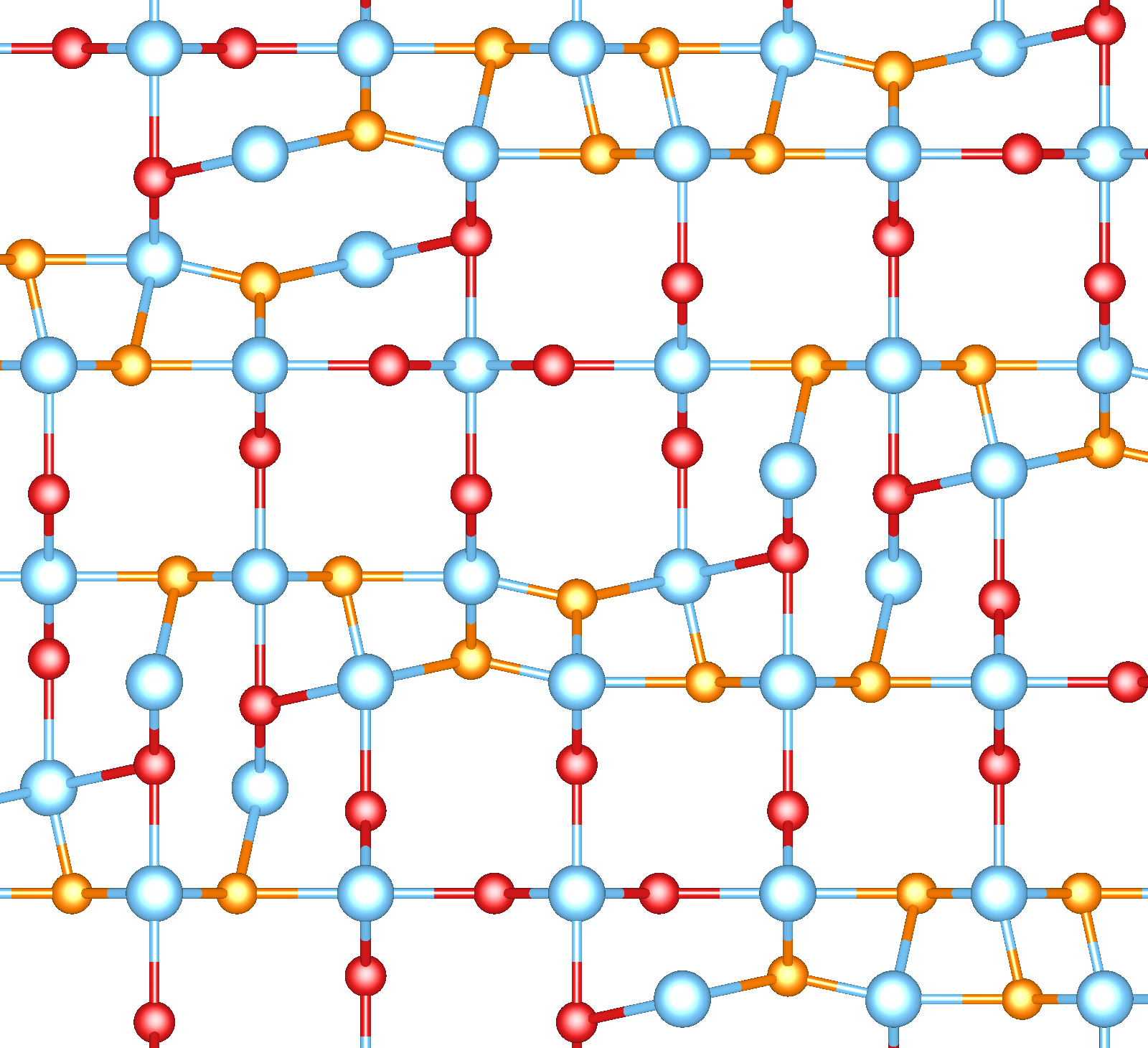}}%
\end{centering}
\caption{\label{fig:magneli_relaxed} (color online)  
  Relaxed Ti$_4$O$_7$ Magn\'eli structure. Overcoordinated O$_{\mathrm{ovc}}$
  atoms (orange) form quasi-one-dimensional channels in [111]. All Ti atoms
  (blue) retain 6-fold coordination. Fully coordinated oxygen atoms appear in
  red. The crystallographic (121) shear plane (see text for affinity to
  rutile) is marked blue.
}
\end{figure*}

\section{\label{sec:results}Results}

\subsection{\label{subsec:energetics}Structural Features and Thermodynamic Stability}


\begin{figure}
\includegraphics[width=6.5cm]{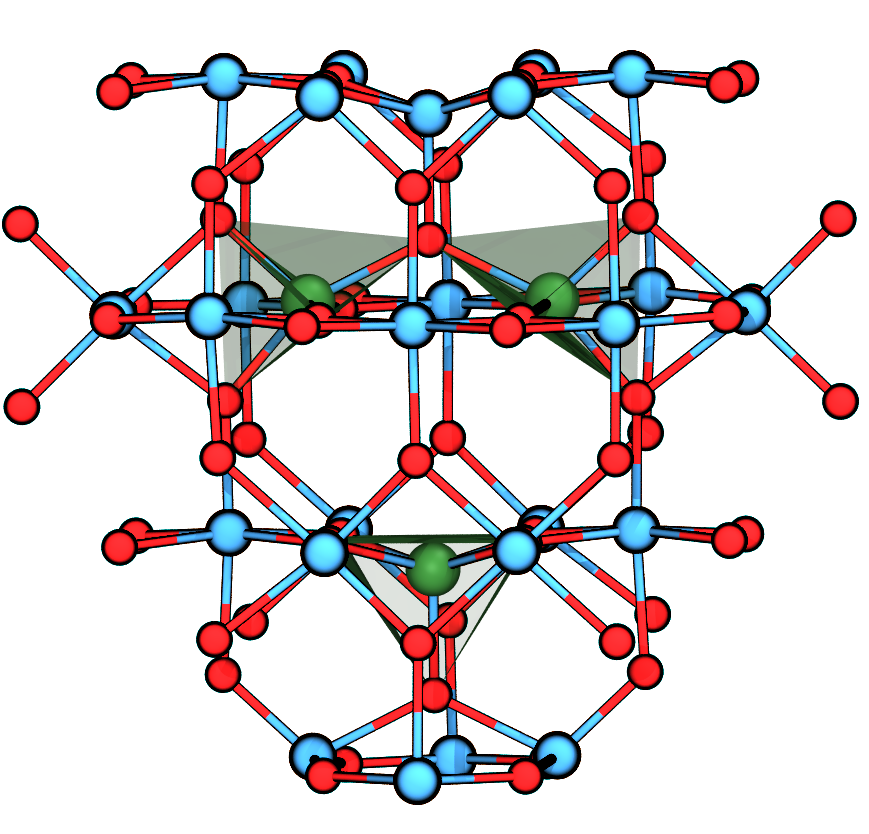}~
\includegraphics[width=1.0cm]{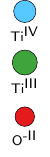}
\caption{\label{fig:iso-vacancy-relaxed} (color online) Relaxed structure of the isolated \VO\ arrangement  
 predicted for 5\,\% \VO\ concentration. The missing oxygen atom induces
 three undercoordinated Ti$^{\mathrm{III}}$ ions, marked in green.}
\end{figure}

Among all structures considered in this study, three classes attract
particular attention due to their energetic appearance in the ground-state
diagram (Fig.~\ref{fig:ceenergies}): 
\begin{enumerate}
  \item The Magn\'eli structures Ti$_5$O$_9$, Ti$_4$O$_7$, Ti$_3$O$_5$, into which none of the structures
    considered in the CE study can relax directly. Having calculated their fully relaxed
    DFT energy separately, they appear to be energetically most
    favorable.
  \item Meta-stable arrangements, having evenly distributed vacant O sites
    separated by at least one occupied site in every direction. These
    structures consistently exhibit formation enthalpies notably above the
    CE ground-state line.
  \item Structures with a certain low-dimensional \VO\ ordering (see below for
    a detailed description), energetically outmatching all other
    \VO\ arrangements covered by the CE study. They are detected at 10\,\%,
    12.5\,\%, and 16.7\,\% \VO\ and form the energetic ground-state line (apart
    from the Magn\'eli structures mentioned above) exhibiting a considerable energy
    gap with regard to all other structures.
\end{enumerate}

Ti$_n$O$_{2n-1}$ Magn\'eli structures are made up of TiO$_2$ rutile blocks,
which are periodically interrupted at (121) shear
planes~\cite{Harada2010,Anderson1967}. The crystallographic operation to move
a rutile block from its perfect rutile position to its Magn\'eli phase
position is (121)$_{\text{rutile}}\frac{1}{2}[0\overline{1}1]_{\text{rutile}}$ relative
to its neighboring rutile block. By deleting overlapping oxygen atoms one
obtains the Ti$_n$O$_{2n-1}$ stoichiometry. The higher the $n$, the larger is
the block of perfect rutile between two shear planes.

Fig.~\ref{fig:magneli_relaxed} shows the relaxed Ti$_4$O$_7$ Magn\'eli
structure, which appears energetically most favorable. As in rutile, all Ti
atoms retain 6-fold coordination. Slightly expanded quasi-one-dimensional
channels of overcoordinated O atoms arrange in [111] direction, separated from
each other in $[0\overline{1}1]$ direction just by two regularly coordinated O atoms.
These approximatively planar arrangements alternate with narrow undisturbed
rutile blocks. While in rutile the Ti octahedra only share edges and vertices,
in the characteristic shear planes of the Magn\'eli phases they share facets.

Due to the substantial restructuring from rutile to a Magn\'eli
phase, as described above, extensive energy barriers are expected to hinder
transformation into the energetically favorable state. None of the \VO\ defect
structures considered in this CE study can relax into it.

Since the cells that are tractable by DFT in large numbers are limited
in size, at low \VO\ concentration the large portions of undisturbed rutile between
separating the \VO\ agglomerates cannot be modeled. Under these circumstances,
the vacant O sites are evenly distributed and separated by at least one 
occupied site in every direction. In this meta-stable arrangement, the vacancy
retains its well known structure surrounded by three truncated TiO octahedra, where the Ti ions are 
regarded as being in the Ti$^{\mathrm{III}}$ oxidation state. 
Exemplarily, Figure~\ref{fig:iso-vacancy-relaxed} shows the relaxed geometry of the lowest energy structure at 5\,\% 
\VO\ concentration. 

For the same reason, higher order
Magn\'eli phases, constructed of \VO\ layers separated by thick layers of undisturbed
rutile, are extremely challenging for computational modeling.
Still, from empirical experience, they are expected to be energetically favorable.

\begin{figure*}
\begin{centering}
\subfloat[\label{fig:layer-prototype}\VO\ arrangement on rutile lattice prior to relaxation]{\includegraphics[width=6.5cm]{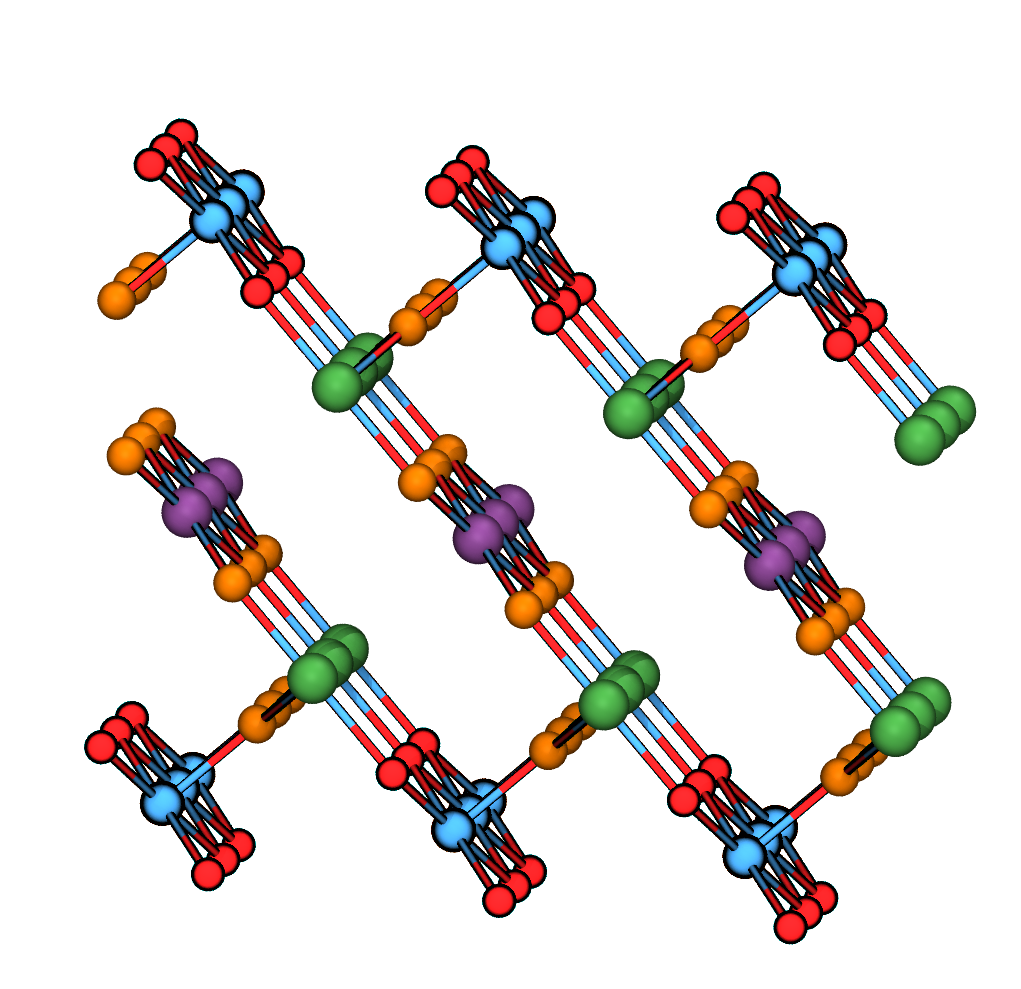}}%
\includegraphics[width=1.0cm]{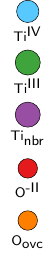}~
\subfloat[\label{fig:1095relax}relaxed structure]{\includegraphics[width=6.5cm]{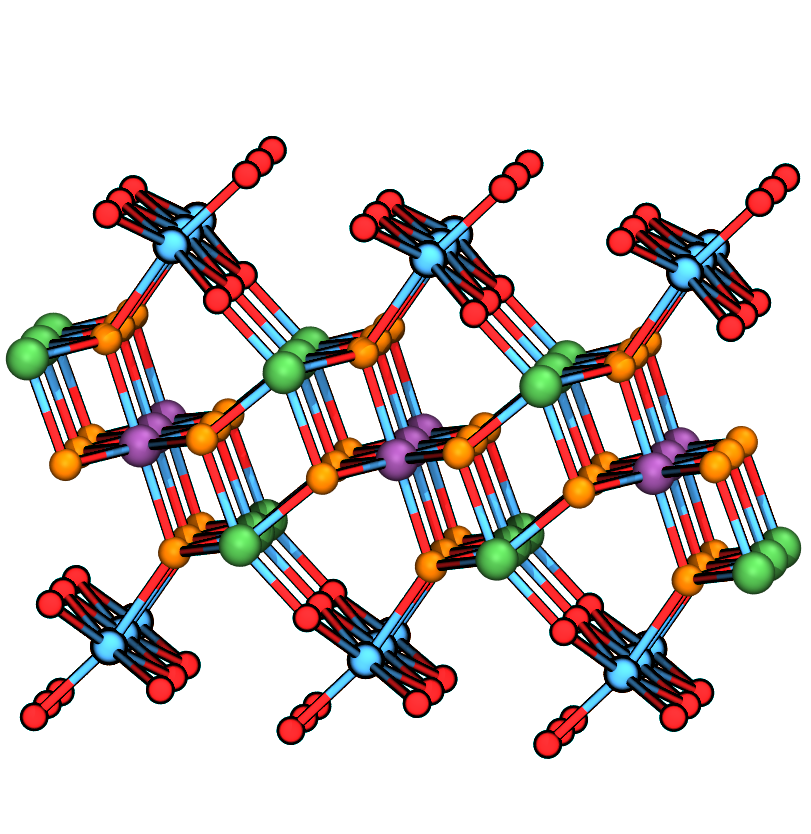}}
\end{centering}
\caption{\label{fig:layeredstructures} (color online)  
Planes of \VO\ relax into Ti$^{\mathrm{III}}$ linked by overcoordinated
O$_{\mathrm{ovc}}$ atoms. For more details and the denotation, see text. }
\end{figure*}

At higher vacancy concentrations,
we find ordering of vacancies in direct neighborhood on a wide variety,
arranged along lines or within slip planes. The prototype of the energetically
favorable defect structure is a \VO\ planar arrangement (see
Figure~\ref{fig:layeredstructures}), detected at 10\,\%, 12.5\,\%, and
16.7\,\% \VO. Prior to relaxation, this structure is
characterized by
two features (Fig.~\ref{fig:layer-prototype}): a (100) \VO\ double plane, (i.e.\ \VO\ 
double rows in [010] and [001] direction,) formed by missing apex oxygen atoms formerly belonging to one 
Ti octahedron, and an odd number of undisturbed Ti octahedra in (100) direction. 
This structure relaxes into the geometry shown in Figure~\ref{fig:1095relax}. It is characterized by 
a layer of five-fold coordinated Ti$^{\mathrm{III}}$ ions, linked by overcoordinated
O$_\mathrm{ovc}$ ions. \footnote{Assigning 
a formal oxidation state to these highly over-coordinated ions based on
their neighbor count yields numbers that bear no connection to the electron
population of the respective atoms. The empirical concept of 'oxidation state' fails in
this context.} The atoms labeled Ti$_\mathrm{nbr}$ are those missing two direct O neighbors in the prototype structure. As
Figure~\ref{fig:1095relax} shows, 6-fold coordination is restored for these
atoms. The layered arrangement of Ti$^{\mathrm{III}}$, Ti$_\mathrm{nbr}$ and
O$_\mathrm{ovc}$ found in the most favorable structures
of our CE separate slightly shifted blocks of perfect rutile from each
other. Thus they are very similar to those
found in the Magn\'eli
phases~\cite{liborio_electronic_2009,liborio_thermodynamics_2008}, lacking the
strong atomic restructuring during formation compared to Magn\'eli phases.


\begin{figure*}
\begin{centering}
\subfloat[\label{fig:712dos} 5\,\%\ \VO\ isolated]{\includegraphics[width=5.5cm]{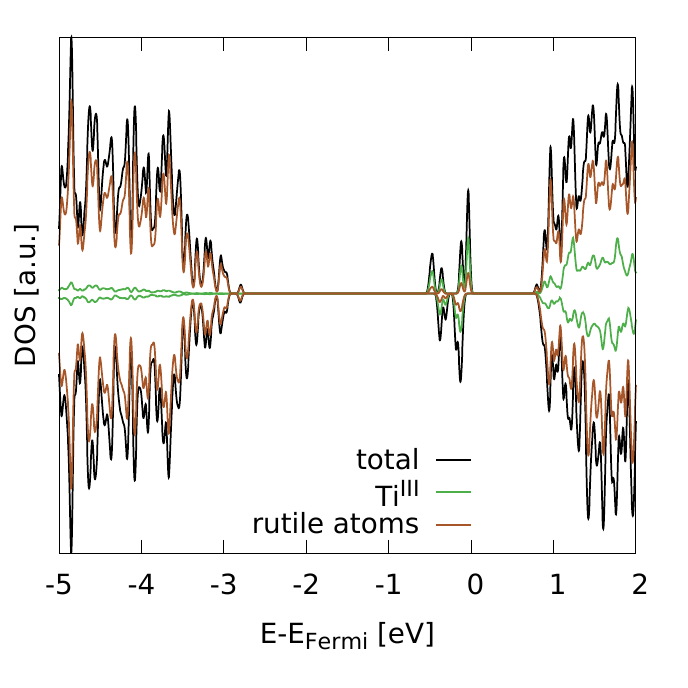}}
\subfloat[\label{fig:magnelidos} Ti$_4$O$_7$ Magn\'eli phase]{\includegraphics[width=5.5cm]{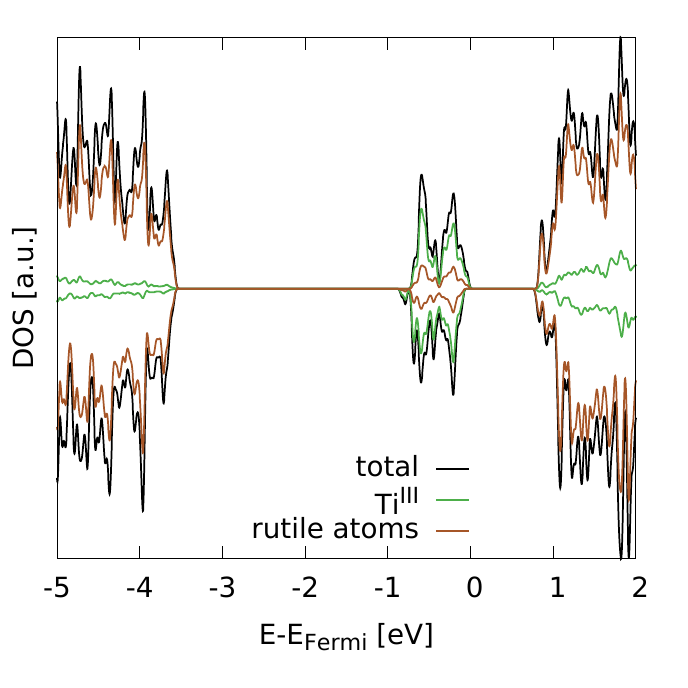}}
\subfloat[\label{fig:1095dos} 16.5\,\%\ \VO\ planar]{\includegraphics[width=5.5cm]{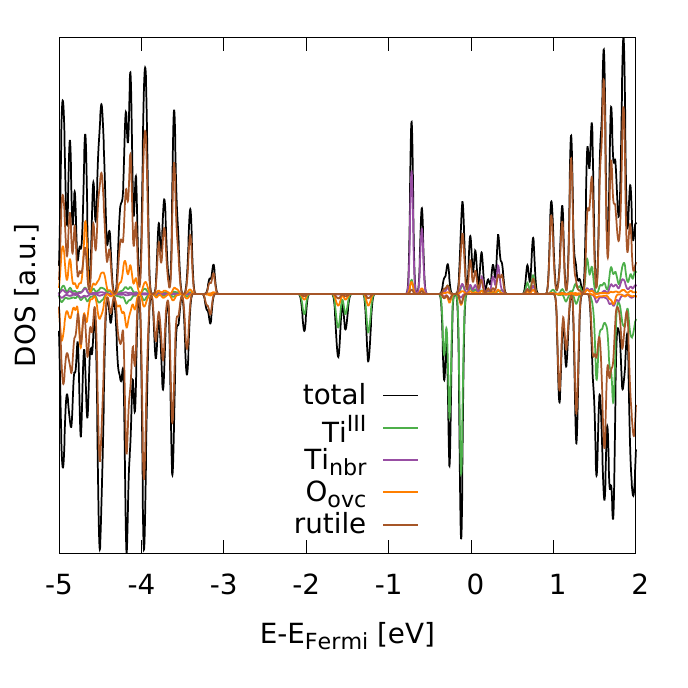}}
\end{centering}
\caption{\label{fig:dosplots} (color online) Spin polarized densities of
  states of isolated \VO\ (cf.\ Fig.~\ref{fig:iso-vacancy-relaxed}), 
  the Magn\'eli phase Ti$_4$O$_7$, and the prototypical planar \VO\ arrangements
  found for 16.7\% \VO\ concentration (cf.\ Fig.~\ref{fig:layeredstructures}).}
\end{figure*}

\subsection{\label{subsec:DOS}Electronic Properties}

Analyzing the densities of states (DOS) of the HSE06 electronic structure calculations performed at the
relaxed geometries, we find a very clear trend illustrated in Figure~\ref{fig:dosplots}: For evenly 
dispersed \VO\ (see e.~g.\ arrangement in Fig.~\ref{fig:iso-vacancy-relaxed}), the electronic structure resembles that of isolated vacancies (cf.\ Fig.~\ref{fig:712dos}). 
For dense \VO\ planes (see e.~g.\ arrangement in Fig.~\ref{fig:layeredstructures}), we note the appearance of a broad, metallic interband close to the
rutile conduction band (Fig.~\ref{fig:1095dos}). This band provides high conductivity even at
low temperatures. Close examination of the projected DOS shows, that the appearance of this
metallic interband is caused by an overlap of defect states originating from Ti$^{\mathrm{III}}$ in the \VO\ layer and states originating from nearby Ti$^{\mathrm{IV}}$.

In contrast, the low temperature form of the 
Ti$_4$O$_7$ Magn\'eli phase exhibits a fully occupied band within the gap (Fig.~\ref{fig:magnelidos}). Our results on the Magn\'eli phase are
in very good agreement with previous calculations by Liborio et~al.~\cite{liborio_electronic_2009} using
DFT+U. Furthermore, they match nicely with the DOS
of Ti$_2$O$_3$ determined by Padilha et~al.~\cite{Padilha2014} using DFT+U and
HSE. Since the latter study contains DOS of higher order
Magn\'eli structures, too, none of them shows a comparable clearly separated,
non-magnetized defect band as the one in Figure~\ref{fig:magnelidos}.

\section{\label{sec:results}Discussion}

Although our findings confirm the initial results of Magyari-K\"ope et~al.~\cite{park_impact_2011}, we interpret our results
with a somewhat different focus. The prohibitive cost of sufficiently precise electronic structure 
calculations severely limited the structural search space of earlier studies and precluded 
cell shape relaxation. Our approach, while still involving enormous
computational effort and being limited to a certain configuration-space
size, allows us to compare all structures of this configuration space instead
of few handpicked samples. Thus, we are able to directly prove that
segregation of \VO\ in rutile TiO$_{2-\delta}$ into planes is favorable over a
uniform distribution.

Even though our present study is static in nature, the very high migration barriers found in bulk
rutile~\cite{knaup_permutation-invariant_2013}, combined with the fact that TiO$_{2-\delta}$ films used
in existing devices are grown in the amorphous state, suggest that the transition is not
\emph{amorphous$\rightarrow$rutile$\rightarrow$Magn\'eli}. We rather expect the vacancy 
agglomeration to occur in a lower density phase, either amorphous or anatase (the 
latter having been observed in situ~\cite{strachan_structural_2009}). This distinction, however small, may prove
technologically very advantageous. It has been shown~\cite{waser_redox-based_2009} that
for commercially viable application in nonvolatile memory, the reaction rate constants 
during switching must be several orders of magnitude larger than during storage or
reading. The phase transition to a rutile based structure in the low resistance state,
which leads to a strong raise in \VO\ migration barriers, facilitates achieving this
difference. To understand, and possibly engineer, the switching process and
storage behavior of memristive devices, it is paramount to first understand
the static, highly stable ON and OFF states. Since we find meta-stable
structures with a clearer metallic nature than Magn\'eli phases, promising
better conductivity, it may be technologically beneficial to seek the
formation of such a non-Magn\'eli structure.

Our approach to the cluster expansion simulation of \VO\ defects in TiO$_2$
is not limited to this specific application, but can be generalized to any 
situation where vacancy or substitutional defects can reside on a distinct
sublattice of a compound material.

The results concerning the \VO\ ordering in TiO$_2$ show a clear tendency for
\VO\ segregation into planes with a very pronounced energy gap with respect to
more disperse distributions. These configurations relax into structures where
planes of Ti$^{\mathrm{III}}$ ions are linked by O$_{\mathrm{ovc}}$
ions. Due to technical constraints, such as limited cell sizes to sustain
DFT manageability and ionic relaxation algorithms not being able
to overcome arising energy barriers, the experimentally found Magn\'eli defect
planes, could not be reproduced in the CE study. Having restored 6-fold
coordination at every Ti atom, the Magn\'eli structures constitute the
ground state---also in this study.

Our detailed analysis based on precise electronic structure calculations shows 
that there is a definite positive correlation between the structural features
of planar Ti$^{\mathrm{III}}$ and O$_{\mathrm{ovc}}$ ions and the formation
of a metallic density of states, as shown in Fig.~\ref{fig:dosplots}. 
Interestingly, the Magn\'eli structure does not exhibit a metallic DOS, but rather
a characteristic deep band of occupied states. The meta-stable \VO\ arrangement shown
in Figure~\ref{fig:1095relax} exhibits a much more pronounced metallic behavior than any Magn\'eli
phase variant examined in our own work as well as earlier DFT+U studies by
Liborio et~al.~\cite{liborio_electronic_2009}.

These findings suggest that the Magn\'eli phase is not necessarily a
prerequisite for the high conductive state of a memristor. Due to the
relatively high \VO\ mobility compared to Ti~\cite{knaup_permutation-invariant_2013}, planar
\VO\ aggregation may happen first, already causing distinct
conductivity. Subsequently, the actual ground state---the Magn\'eli
phase---may arise by shifting the rutile layers among the \VO\ planes and
finally form the long-lasting low-resistance state.

Further, the results cast some doubt upon the identification of the crystal structure of the
conductive filaments in on-state memristor devices as Magn\'eli phases. The relaxed planar
\VO\ arrangements are structurally very similar to the Magn\'eli phase. Specifically,
both phases exhibit the same arrangement of disturbed planes at the same inter-plane
distances. This makes the distinction between the Magn\'eli phase and the 
planar \VO\ arrangements found in our CE calculations extremely challenging.

\section{\label{sec:conclusion}Conclusion}

In summary, we performed a broad and systematic study of the thermodynamics of
 \VO\ arrangements in rutile TiO$_2$, combined with highly accurate electronic 
 structure analysis of the most favorable structures. 
Our results are in accordance with previous experimental reports indicating a
preferential \VO\ segregation into planar arrangements~\cite{kwon_atomic_2010}, and confirm the
earlier hypothesis that these arrangements cause the observed
high conductivity~\cite{park_impact_2011}.
 The Magn\'eli phase does not seem
 to be a mandatory prerequisite for the low resistive state of memristors. 
 Some contradictions between the 
 calculated electronic structure of the Magn\'eli phase and observed metallic behavior
 point to two open fundamental questions: On the theoretical side, precise and detailed band structure
 calculations of the Magn\'eli phase are needed,
 while experimentally it must be endeavored 
 to distinguish between the Magn\'eli phase and other planar \VO\ arrangements.

\begin{acknowledgments}
We gratefully acknowledge financial support from the German Research Foundation (DFG) via SFB~986~"M$^3$", project~A4.
J.M.K.\ is grateful for a grant in the "own position" program by the German Research Association (DFG).
Atomic structure representations were created using \textsc{Vesta}~\cite{MommaVesta2011} and VMD~\cite{humphrey_vmd_1996}. CE was performed using the
\textsc{Uncle} code~\footnote{\textsc{Uncle} is licensed by Materials Design. \texttt{info@materialsdesign.com}}.
The authors thank Dr.\ Peter De\'ak for intensive discussion about properties and simulation of
\VO\ defects in TiO$_2$.
\end{acknowledgments}

\end{document}